\documentclass{emulateapj}
\usepackage{color, hyperref, epsfig}
\usepackage{apjfonts, natbib}

\newcommand{\hst}{\emph{HST}}
\newcommand{\xmm}{\emph{XMM-Newton}}
\newcommand{\ha}{H\ensuremath{\alpha}}

\newcommand{\mgii}{Mg\,{\footnotesize II}}
\newcommand{\feii}{Fe\,{\footnotesize II}}

\newcommand{\lum}{erg~s\ensuremath{^{-1}}}
\newcommand{\lbol}{\ensuremath{L\mathrm{_{bol}}}}
\newcommand{\ledd}{\ensuremath{L\mathrm{_{edd}}}}
\newcommand{\msun}{\ensuremath{M_{\odot}}}
\newcommand{\kms}{\ensuremath{\mathrm{km~s^{-1}}}}
\newcommand{\mbh}{\ensuremath{M_\mathrm{BH}}}
\newcommand{\wise}{\emph{WISE}}
\newcommand{\neowise}{\emph{NEOWISE}}
\newcommand{\tsub}{\ensuremath{T_\mathrm{sub}}}
\newcommand{\tdust}{\ensuremath{T_\mathrm{dust}}}
\newcommand{\rsub}{\ensuremath{R_\mathrm{sub}}}
\begin{document}

\title{
Mid-infrared flare of TDE candidate PS16dtm: Dust echo and implications for the spectral evolution
}
\author{
%\emph{Tentatively:}
Ning~Jiang\altaffilmark{1,2},
Tinggui~Wang\altaffilmark{1,2},
Lin Yan\altaffilmark{3,4}, 
Ting~Xiao\altaffilmark{5},
Chenwei~Yang\altaffilmark{1,2}, 
Liming~Dou\altaffilmark{6},
Huiyuan~Wang\altaffilmark{1,2}, 
Roc~Cutri\altaffilmark{4},
Amy~Mainzer\altaffilmark{7}}
\altaffiltext{1}{Key laboratory for Research in Galaxies and Cosmology,
Department of Astronomy, University of Science and Technology of China,
Chinese Academy of Sciences, Hefei, Anhui 230026, China; ~jnac@ustc.edu.cn}
\altaffiltext{2}{School of Astronomy and Space Sciences, 
University of Science and Technology of China, Hefei, Anhui 230026; ~twang@ustc.edu.cn}
\altaffiltext{3}{Caltech Optical Observatories, Cahill Center for Astronomy and Astrophysics,
California Institute of Technology, Pasadena, CA 91125, USA; lyan@ipac.caltech.edu}
\altaffiltext{4}{IPAC, Mail Code 100-22, California Institute of Technology, 1200 E. California Boulevard, Pasadena, CA 91125, USA}
\altaffiltext{5}{
Key Laboratory for Research in Galaxies and Cosmology,
Shanghai Astronomical Observatory, Chinese Astronomical Society,
80 Nandan Road, Shanghai 200030, China}
\altaffiltext{6}{Center for Astrophysics, Guangzhou University, Guangzhou 510006, China}
\altaffiltext{7}{Jet Propulsion Laboratory, California Institute of Technology,
Pasadena, CA 91109, USA}
%%%%%%%%%%%%%%%%%%%%%%%%%%%%%%%%%%%%%%%%%%%%%%%%%%%%%%%%%%%%%%%%%%%%%%%%%%%%%%%%

\begin{abstract}

PS16dtm was classified as a candidate tidal disruption event 
(TDE) in a dwarf Seyfert 1 galaxy with low-mass
black hole ($\sim10^6M\odot$) and has presented various 
intriguing photometric and spectra characteristics. Using the
archival \wise\ and the newly released \neowise\ data, we 
found PS16dtm is experiencing a mid-infrared (MIR) flare
which started $\sim11$ days before the first optical detection.
Interpreting the MIR flare as a dust echo requires close pre-existing dust with 
a high covering factor, and suggests the optical flare may have brightened slowly 
for some time before it became bright detectable from the ground.
More evidence is given at the later epochs. At the peak of the 
optical light curve, the new inner radius of the dust torus has grown 
to much larger size, a factor of 7 of the initial radius due to 
strong radiation field. At $\sim150$\,days after the first optical detection, 
the dust temperature has dropped well below the sublimation temperature. 
Other peculiar spectral features shown by PS16dtm are the transient, prominent
FeII emission lines and outflows indicated by broad absorption lines 
detected during the optical flare. Our model explains the enhanced \feii\
emission from iron newly released from the evaporated dust. 
The observed broad absorption line outflow could be explained 
by accelerated gas in the dust torus due to the radiation pressure. 

\end{abstract}

\keywords{galaxies: individual (PS16dtm) --- galaxies: active --- galaxies: nuclei}

\section{Introduction}

When a star passes within the tidal radius of a supermassive black hole (SMBH),
it will be tidally disrupted and partially accreted by the
central black hole, producing luminous flares in the soft X-ray,
UV and optical wavelengths. The flare usually lasts a few months to years
(decaying as $t^{-5/3}$) and releases a total energy of $10^{53}$~erg for
a solar-type star (see Rees 1988; Phinney 1989).
This phenomenon is known as tidal disruption event (TDE).
Despite that TDE is a ubiquitous consequence of the existence of SMBHs
at the cores of galaxies, yet almost all known TDEs are discovered in dormant
galaxies but not in active galactic nuclei (AGN).
In fact, there's no physical process specifically forbids disruptions from occurring and 
the rates are expected to be even higher when the black holes are 
originally active (Hills 1975; Karas \& Subr 2007).
The problem is that it's much more challenging to identify TDEs in AGNs.
First, the TDE flare manifests itself as a much lower contrast signal
relative to the permanently bright accretion disk, while the discoveries of TDEs are usually
triggered by a sudden and intense brightening of the galaxy centers in time-domain surveys.
Second and the most intractable problem is one can barely connect any
particular flare in a classical AGN with a TDE, because the long-lived accretion
itself can potentially cause high-amplitude variability,
by some special instability of the accretion disk.
Given the number of transients now discovered by time domain surveys, and the 
limited resources for spectroscopic follow-up, transients coincident with the centers 
of active galaxies have often been attributed solely to AGN variability.
Observationally, quite a few studies of the outburst signatures in Seyfert galaxies 
have considered the possibility of TDEs in AGNs 
(e.g., Peterson \& Ferland 1986; Brandt et al 1995),
while none of them can draw convincing conclusion due to the potential intrinsic 
variability (see section 7 in Komossa 2015).

Recently, Blanchard et al (2017) have claimed that they have discovered a TDE candidate 
PS16dtm in SDSS~J015804.75-005221.8 (hereafter J0158 for short), 
which is a Seyfert~1 galaxy with low-mass black hole 
($\sim10^6$~\msun, Greene \& Ho 2007; Xiao et al. 2011). 
This event shows several distinctive characteristics revealed by
multi-wavelength photometric and spectroscopic follow-up monitorings:
(1) the light curve has brightened by about two magnitudes 
in 50 days relative to the archival host brightness and then exhibited a plateau phase for
about 100 days.
(2) it is undetected in the X-rays by Swift/XRT to a limit an order of magnitude 
below an archival \xmm\ detection.
(3) the spectra, spanning UV to near-IR, exhibit multicomponent hydrogen emission lines 
and strong \feii\ emission complexes, show little evolution with time, and
closely resemble the spectra of NLS1s while being different from those of Type~IIn SNe.
(4) the UV spectra shows a broad \mgii\ absorption component, indicative of an outflow.
These features make it quite distinctive from known TDEs found in normal galaxies
and may provide us an unique opportunity to understand the physical mechanisms 
associated with both TDEs and AGNs, such as the arising of \feii\ lines in NLS1s
and the driven mechanism of broad absorption line (BAL) quasars.

In this paper, we report a mid-infrared (MIR) flare detected in the location coincident 
with PS16dtm using data from the $\emph{Wide-field Infrared Survey Explorer}$ 
(\wise; Wright et al. 2010; Mainzer et al. 2014).
Recent detections of MIR flares in TDEs are all explained successfully as the dust echoes 
(Jiang et al. 2016; Dou et al. 2016, 2017; van Velzen et al. 2016).
The MIR echoes can help us probe the dusty medium within a few parsecs
surrounding SMBHs and measure the total energy released during the event.
The MIR flare found in PS16dtm may also give us useful insights to 
unveil its nature and explain its various interesting properties mentioned above. 
We assume a cosmology with $H_{0} =70$ km~s$^{-1}$~Mpc$^{-1}$, $\Omega_{m} = 
0.3$, and $\Omega_{\Lambda} = 0.7$.  At a redshift of $z=0.08044$, PS16dtm
(associated with J0158) has a luminosity distance of 365.4~Mpc.

\section{Mid-infrared and optical light curves}

The \wise\ has performed a full-sky imaging survey in four broad mid-infrared
bandpass filters centered at 3.4, 4.6, 12 and 22~$\mu$m (labeled W1-W4)
from 2010 February to August.
The solid hydrogen cryogen used to cool the W3 and W4 instrumentation
was depleted later and it was placed in hibernation in 2011 February.
\wise\ was reactivated and renamed \neowise-R since 2013 October, using only W1 and W2,
to hunt for asteroids that could pose as impact hazard to the Earth.
\wise\ has scanned a specific sky area every half year and thus yielded 9-11 times
of observations for every object up to now.
For each epoch, there are 10-20 single exposures.  
We use only the best quality single frame images by selecting only detections
with data quality ag 'qual\_frame'>0.

\begin{figure}
\centering
\includegraphics[width=9cm]{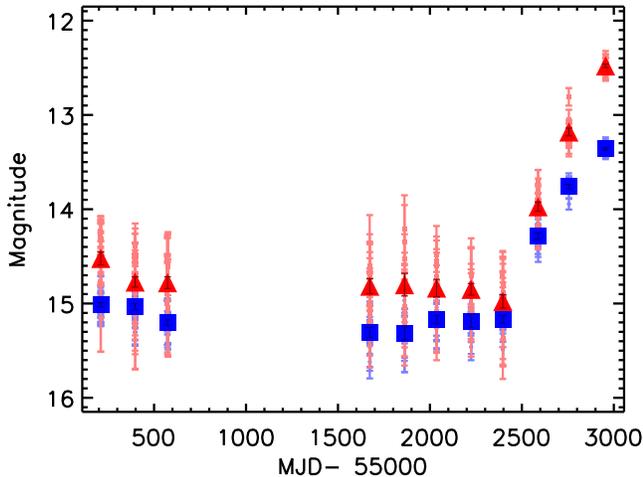}
\caption{
The \wise\ light curve of PS16dtm.
Blue squares: W1($3.4\mu$m) band. Red triangles: W2($4.6\mu$m).
}
\label{wiselc}
\end{figure}

J0158 is a point-like source in all of the \wise\ images and there's no any
potential contamination within 20\arcsec. 
We have collected all of its \wise\ photometric data, including from ALLWISE and
\neowise-R catalogs. The data are distributed in 11 epochs at intervals of about 
six months up to 2017 July. 
The light curves in W1 and W2 are presented in Figure~\ref{wiselc}.
First, no reliable variability is detected within each epoch and thus
we then average these data to obtain a mean value at each epoch following our
previous works (Jiang et al. 2012; 2016). 
Second, there's no significant variability in the first 8 epochs;
possible slight variation can be due to the usual AGN fluctuation.
We take these 8 epochs as the quiescent state, in which the MIR emission 
can be considered as the background emission from the host galaxy and AGN, contributing 
an averaged magnitude of $15.19\pm0.04$ and $14.82\pm0.05$ at W1 and W2, respectively.
However, it has become $\sim1$ mag brighter in both bands from the ninth epoch,
that is MJD=57,587 or 2016 July 18. The flux get even more higher in the next
two epochs, suggesting it is a MIR flare. We have presented the background-subtracted 
light curves in the top panel of Figure~\ref{lc} 
and denoted the three brightened epochs as "E1", "E2" and "E3".

\begin{figure}
\centering
\includegraphics[width=9cm]{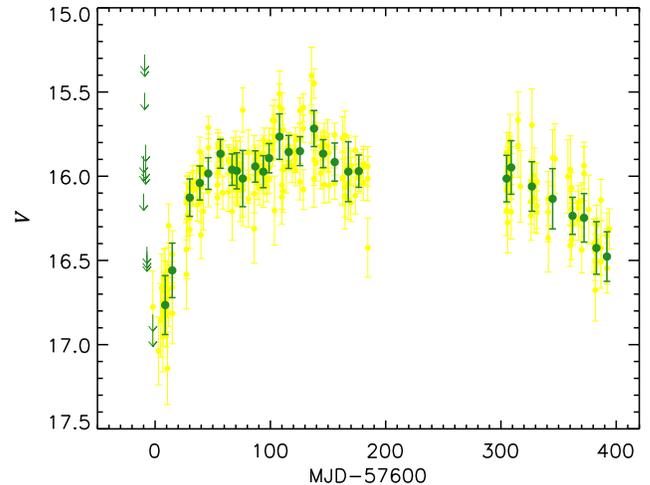}
\caption{
The ASAS-SN $V$-band light curves of PS16dtm (yellow), among which the data points
before MJD$\sim57,600$ are only upper limits impact by the bright moon phase.
The binned data (every 9 points) are plotted in dark green.
}
\label{vlc}
\end{figure}

We have noticed that as metioned in Blanchard et al. (2017), PS16dtm was discovered 
by Pan-STARRS Survey for Transients (PSST; Huber et al. 2015) on 2016 Aug 12 (MJD=57,612) 
and by the All-Sky Automated Survey for Supernovae (ASAS-SN; Shappee et al. 2014) on 2016
July 29 (MJD=57,598), that is 25 and 11 days later than "E1", respectively.  
In order to check and update its optical light curve, we have retrieved the 
$V$-band photometry up to now from ASAS-SN light curve server 
(Kochanek et al. 2017)~\footnote{https://asas-sn.osu.edu/} and presented them in 
Figure~\ref{vlc}.
PS16dtm has risen to its optical peak within $\sim50$ days since its detection and then
kept at an almost constant level for $\sim100$ days.  
After $\sim100$ more days, it enters into the visibility window of ASAS-SN again and 
shows a slow declining trend, matching with a TDE scenario.

As J0158 is undetected by ASAS-SN before the outburst, we turn to an estimation of
the quiescent $V$ magnitude from the SDSS photometry taken at more than 
a decade before (MJD=52,224).
We adopted the transformation ($V$=g-0.52*(g-r)-0.03) introduced in Jester et al. (2015), 
yielding $V=18.31$ with an error of 0.05, that is indeed far below the ASAS-SN detection 
limit ($V\sim17$).
The background-subtracted $V$-band light curves are shown in Figure~\ref{lc}.

\section{Analyses and Results}

\subsection{The early detection of the MIR flare: not inconsistent with a dust echo}

\begin{figure}
\centering
\includegraphics[width=8cm]{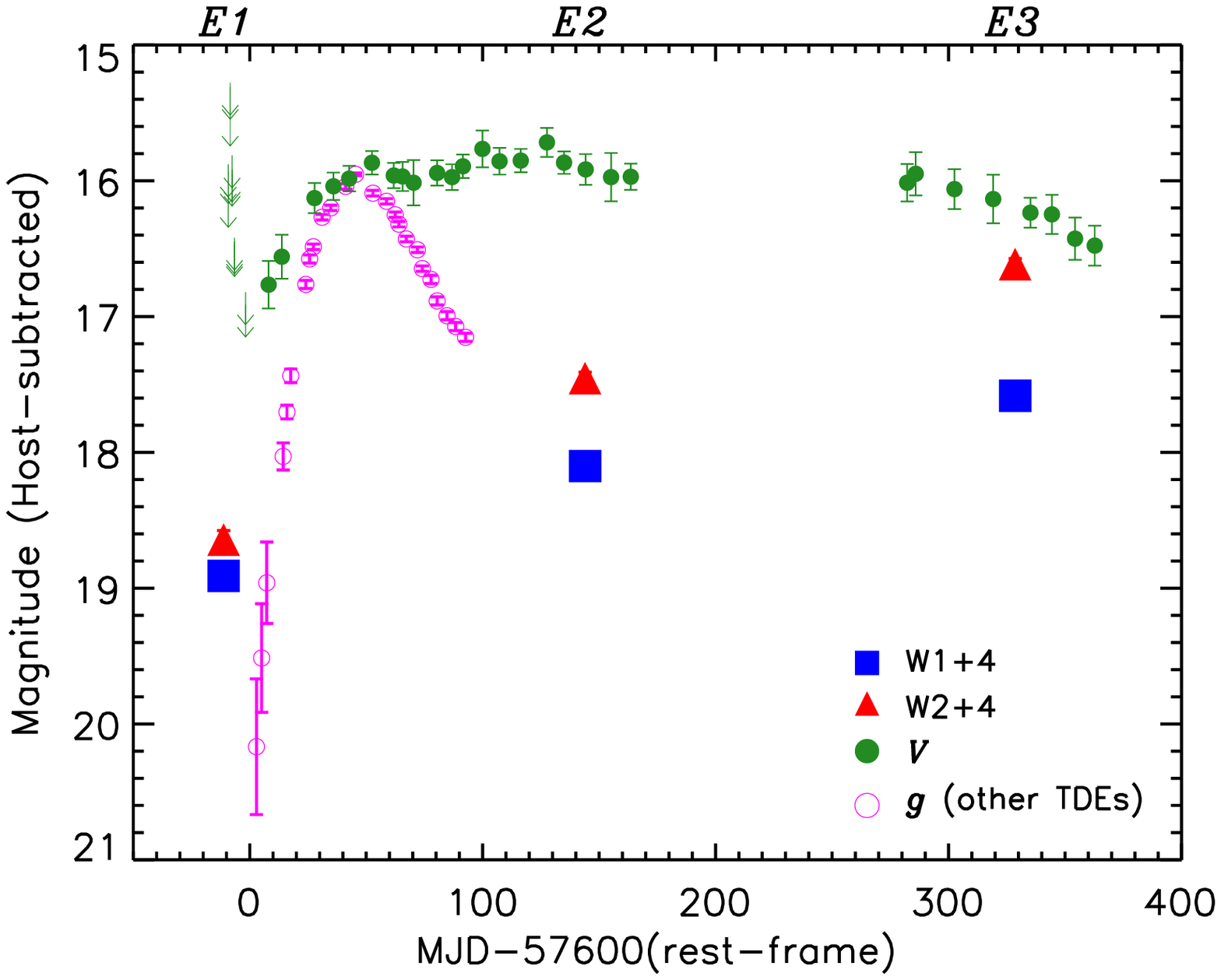}
\includegraphics[width=8cm]{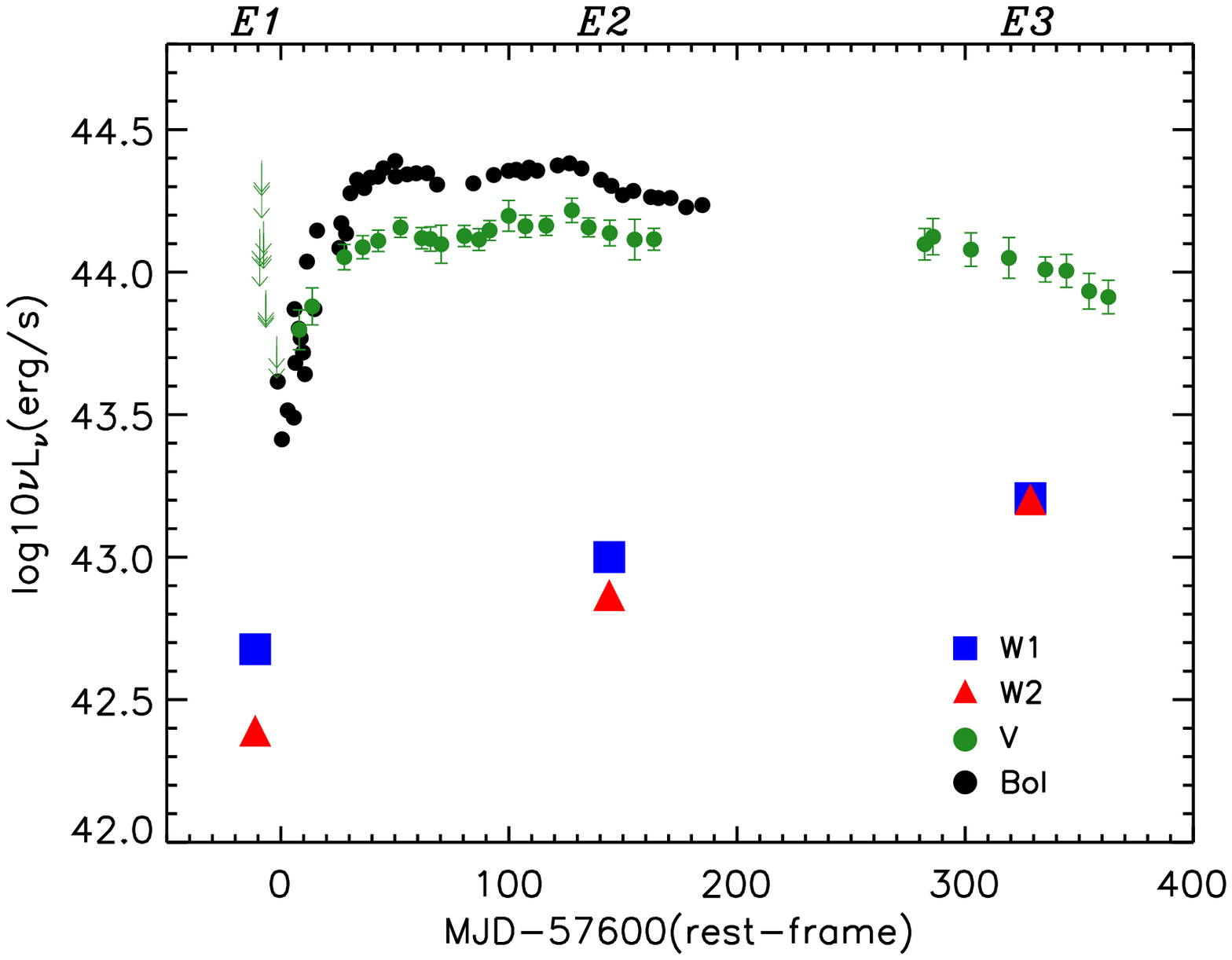}
\caption{
The \wise\ and $V$-band background-subtracted light curves (top) and luminosity evolution
(bottom) of PS16dtm.
The black dots are optical bolometric luminosity taken from the
top panel of Fig.6 of Blanchard et al. (2017).
In the top panel, $g$-band light curves (magenta circles) of two other well-known TDEs
(PTF09ge and PS1-10jh, their light curves are almost identical and thus we plot them
as a whole, see also Fig. 11 of Arcavi et al. 2014) have also been
overplotted and they have been scaled to the same peak time and magnitude of PS16dtm
for ease of comparison.
}
\label{lc}
\end{figure}

Previous detections of MIR flares in TDEs are all attributed to dust echoes 
in view of the basic fact that the MIR flares lag behind the optical
(Jiang et al. 2016; Dou et al. 2016, 2017; Van velzen et al. 2016).
Nevertheless, the \wise\ detection time of the PS16dtm MIR flare is suprisingly 
earlier than the optical detection.
At the face value, it is hard to explain the MIR flare of PS16dtm as a dust echo as usual.
However, the particularities of the event must be taken into consideration to see whether
the contradiction is explicable or not.

First, the host of PS16dtm before the outburst is a low-luminosity Seyfert~1 galaxy with 
$L_{\rm X}=1.2\times10^{42}$~\lum\ (Pons \& Watson 2014) and
$\nu L_{\nu}$(5100\AA)/$L_{\rm X}$=0.5 (Blanchard et al. 2017).
The bolometric luminosity (\lbol) can be roughly estimated as
$\lbol=9\times\nu L_{\nu}$(5100\AA)=$5.4\times10^{42}$~\lum\ (Kaspi et al. 2000),
that's $\sim0.04~\ledd$ assuming a BH mass $\sim10^6$~\msun\ (Xiao et al. 2011).
At this quiescent state, the inner radius of dusty torus can be estimated from
the sublimation radius (e.g. Namekata \& Umemura 2016).
For a typical AGN spectrum, the dust sublimation radius is given by
\begin{eqnarray}
\rsub & = &  0.121~pc \left(\frac{\lbol}{10^{45}\;\mathrm{erg\;s^{-1}}}\right)^{0.5}\left(\frac{\tsub}{1800\;\mathrm{K}}\right)^{-2.804} \left(\frac{a}{0.1\;\micron}\right)^{-0.510}, \label{eq:Rsubiso}
\end{eqnarray}
where $\lbol$ is the bolometric luminosity of AGN, $\tsub$ is the sublimation temperature 
of dust grain, and $a$ is the grain radius.
On the other hand, the torus inner radius can be estimated from the $K$-band reverberation 
mapping result (e.g. Suganuma et al. 2006; Kishimoto et al. 2007). For example,
Kishimoto et al (2007) has given a time-lag radii $R_{\rm \tau K}$ as function of
$V$-band luminosity by fitting data points from Suganuma et al (2006):
\begin{equation}
  R_{\rm \tau K}=0.47\left(\frac{6\nu L_{\nu}(V)}{10^{46}~\rm erg/s}\right)^{1/2}
\end{equation}
Both methods yield a physical scale of $\sim9\times10^{-3}pc$ or 10 light days for the 
BH in J0158.
The small distance of the dusty torus to the continuum source may explain 
the early detection of the MIR flare because the time-delay is too short to measure.
Naively assuming that the increased MIR flux is coming from the dust thermal emission
heated by the TDE, the derived blackbody temperature is presented in Figure~\ref{temp}.
The blue W1-W2 color (0.27 magnitudes) at E1 corresponds to
a blackbody temperature of $\sim2.5\times10^3$~K, supporting that
the dust responsible for the early MIR flare is located at around sublimation radius.
Second, at the early rising phase, the MIR luminosity is $\gtrsim10\%$ of the
bolometric one (see bottom panel of Figure~\ref{lc}), indicating that the 
covering factor of the dust is fairly high. 
In this scenario, the surrounding dust may absorb a significant fraction of 
UV-optical photons and then re-emit in the IR, attenuating the optical flare but 
boosting the MIR flare.
Third, it's difficult to detect TDEs in active galaxies in the early rising,
prevented by the dilution of the background emission, including the intrinsic BH accretion 
and instability. In contrast, the MIR emission is dominated by hot dust heated 
by black hole accretion and much less diluted by the host galaxy.
Moreover, the photometry basing on ground-based optical survey is severely 
limited by sky conditions, such as the moon phase impact on 
PS16dtm before its detection by ASAS-SN.

Above three, the puzzling fact that the flare was detected in MIR earlier 
than optical is unusual but not unreasonable. 
The optical flare should have been rising for some time before it became bright
enough for ASAS-SN to detect. If that is the case, the rising timescale of this event is 
probably longer than other TDEs. It's interesting to compare the 
optical light curves of PS16dtm with two other well-known TDEs monitored since
early rising stage (PTF09ge and PS1-10jh) shown in Arcavi et al. (2014). 
Obviously, the rising slope of PS16dtm is much smoother, agreeing with our expectation. 
Further studies of the physical nature behind different rising behaviours is valuable
to explore when we have more well-sampled TDE light curves in the future.

This problem can be also reconciled in alternative ways under the TDE scenario. 
According to the calculation of Guillochon \& Ramirez-Ruiz (2015), there is a delay 
between the flare and the disruption of the star of order years for BH mass of $10^7$~\msun.
With such delay, the unbound debris can reach a distance of a light month when we
see optical flare. If the unbound debris collides with molecular clouds in dust torus,
it will produce strong shocks and emit UV/X-rays, which can heat the surrounding dust.
If a substantial fraction of kinetic energy of the debris is converted to the
radiation energy, it may explain the early emission of the observed MIR flare.
The other interpretation has to do with the uncertain origin of optical emission in 
TDE disks.  Naive models of TDE disks predict that most emission should be in the 
soft X-ray/EUV, and the optical emission should be orders of magnitude dimmer than 
what we have observed for optically discovered TDEs (Miller 2015).  
The most common solution to this puzzle is that TDEs create some 
kind of "reprocessing layer" at $\sim100-1000$ times of Schwarzschild radii to 
absorb and downgrade the hard photons from the central engine.  
This layer could either be a quasi-static bound envelope of poorly circularized 
debris (Metzger \& Stone 2016) or some sort of outflowing wind (Guillochon et al. 2014).
It is possible that this reprocessing layer takes some time to set up, and
thus the optical emission lags behind the MIR emission heated by the  
bare, unreprocessed X-ray/EUV emission from the inner accretion disk. 
These scenarios are interesting and possible, yet current data for PS16dtm are
insufficient to test and can only be checked by other X-ray discovered TDEs.

\subsection{Further evidence for the MIR flare as a dust echo}

As the bolometric luminosity of PS16dtm increases, dust in the inner region
evaporates gradually, pushing the inner side of dusty torus outward.
The peak luminosity of the flare ($\sim2.2\times10^{44}$~\lum) determines
a new torus inner radius of about $\sim7$ times larger than its quiescent size,
or 70 light days ($\rsub\propto\lbol^{1/2}$). 
During this process, the time lag effect
on the light curve becomes important because the UV continuum rising timescale
(of order 30 days) is shorter than the light crossing time scale.
Basing on this evolution picture, more evidence for the dust echo appears at 
the second and third \wise\ detection epoch (E2 and E3), that's rest-frame $\sim150$
and 330 days after the optical detection.

\begin{figure}
\centering
\includegraphics[width=9cm]{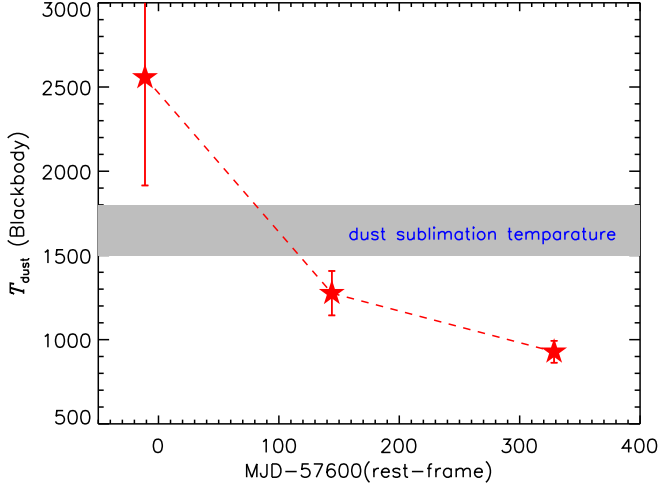}
\caption{
The evolution of dust blackbody temperature fitted by the background-subtracted \wise\ data.
The grey region denotes the dust sublimation temperature range.
}
\label{temp}
\end{figure}

The bolometric luminosity at E2 remains nearly the same as the peak, while the MIR color 
becomes much redder (W1-W2=0.65) than E1 (see Figure~\ref{lc}),
indicating a blackbody dust temperature of $1.3\times10^3$~K.
This fact implies that the average dust temperature (\tdust) may have dropped to 
well below the sublimation temperature (see Figure~\ref{temp}); 
the dominated IR emission comes from further away, 
and the dust around the inner region must be optically thin.
It's also notable that the IR luminosity is only a small fraction ($<0.1$)
of bolometric luminosity at E2, implying a smaller covering factor at E2 than E1
as a result of dust sublimation.
Since the optical light curve is almost flat and declining very slowly at E3,
the dust emission from inner region should not change much. However, the dust emission
located more further out is add-on and thus it will lead to a continuously rising 
light curve yet a declining dust temperature in the case of extended optically 
thin dust. The prediction is exactly consistent with the \wise\ observation 
at E3, which has a higher MIR liminosity yet redder color (W1-W2=0.97), namely 
lower dust temperature ($\sim900$~K).

\begin{figure}
\centering
\includegraphics[width=8cm]{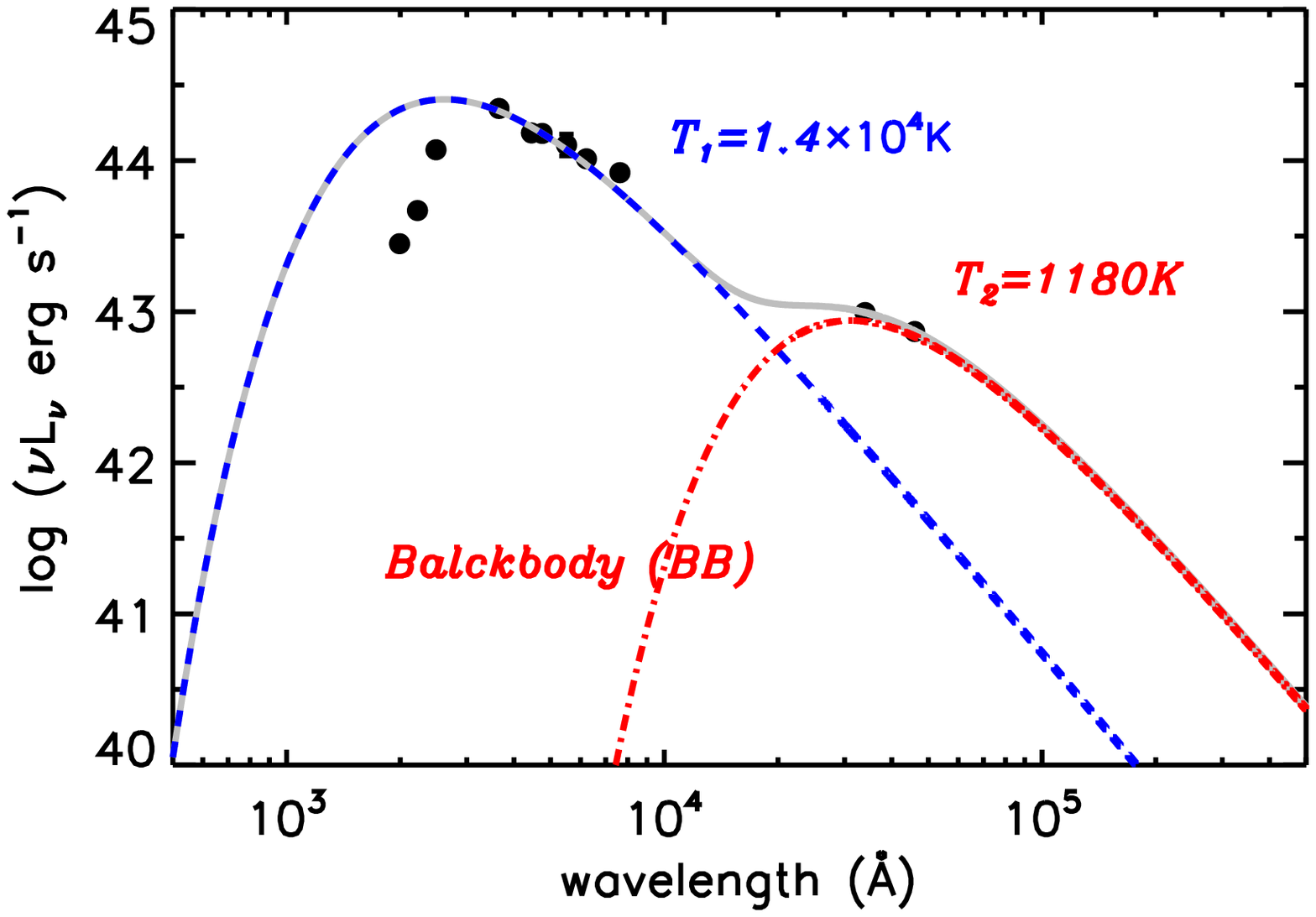}
\includegraphics[width=8cm]{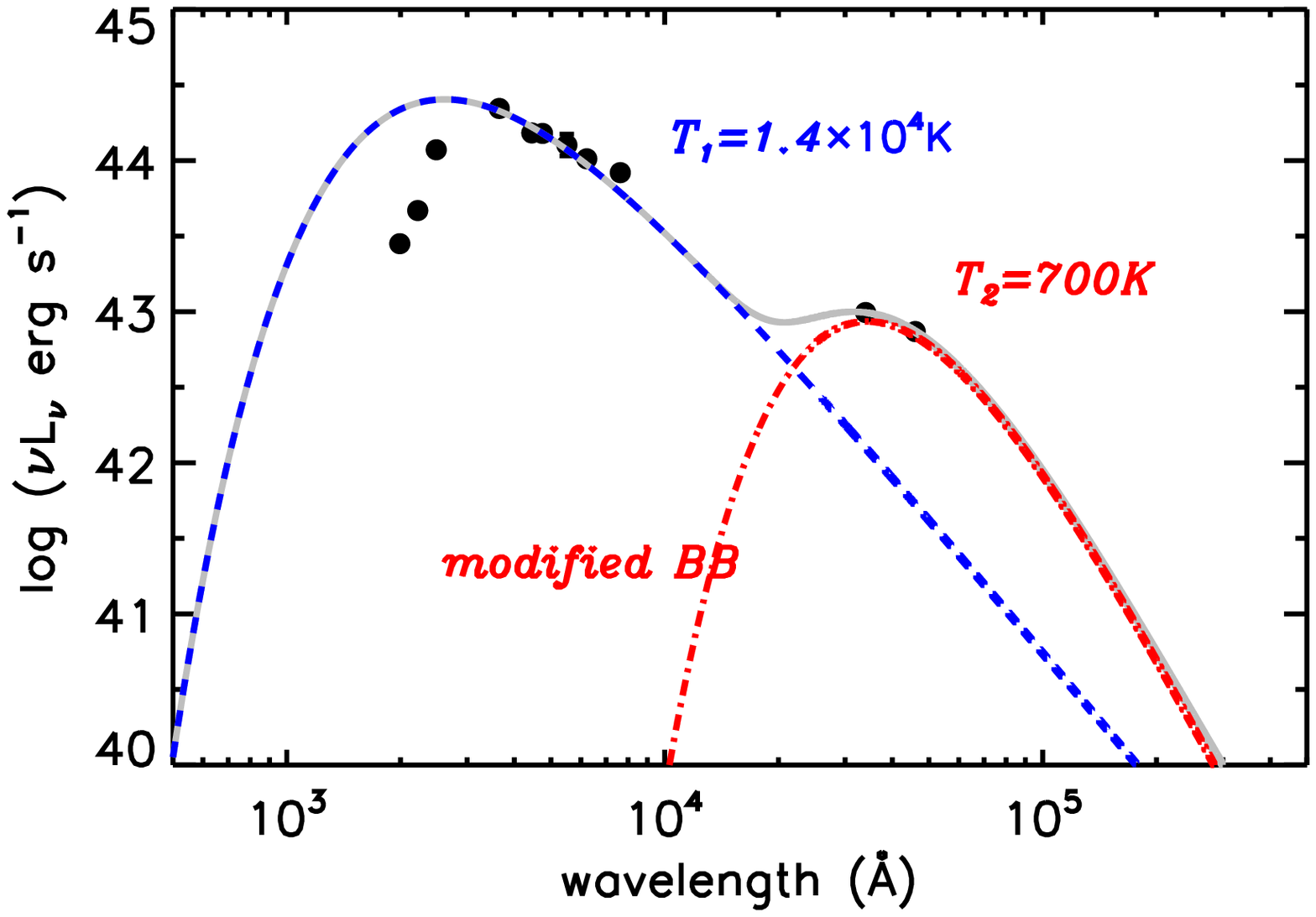}
\caption{
The background-subtracted SED of PS16dtm at E2 (MJD$\approx$57,755).
The UV-optical photometry is drawn from Table~1 of Blanchard et al. (2017).
We have tried to fit the SED using double-blackbody model:
the one dominating UV-optical emission is fixed to a blackbody temperature of
$1.4\times10^4$~K (blue dashed line) as suggested in Blanchard et al. (2017);
the other one (red dotted-dashed line) is primarily from the dust emission
assuming a simple blackbody (top panel) or considering the dust absorption efficiency
(bottom panel).
}
\label{sed}
\end{figure}

To obtain a stricter \tdust\ value and more quantitive calculation, 
we have collected the UV-optical-IR photometric 
data observed quasi-simultaneous with E2 (see Figure~\ref{sed}).
First, we tried to fit the spectral energy distribution (SED) with 
double-blackbody model, yielding a \tdust\ of $1.2\times10^3$~K.
During the fit, we have fixed the blackbody temperature responsible for the
UV-optical emission to $1.4\times10^4$~K as given in Blanchard et al. (2017).
Assuming the equilibrium temperature \tdust\ is determined by the
balance between the radiative heating by the absorption of UV-optical photons
and the thermal re-emission in the IR:
\begin{equation}
  \frac{L_{\rm bol}}{4\pi R^2} \pi a^2  = 4\pi a^2 \sigma T^4.
\end{equation}
The calculated \lbol\ is
$5.2\times10^{44}$~\lum, comparable yet slightly higher than the observed peak \lbol.
Here we have used the dust distance calculated from a light travel time of 120 days,
that is the time interval since the peak.
Similar to the case of ASASSN-14li (Jiang et al. 2016),
the blackbody assumption for the dust emission could be over-simplistic and the
dust absorption efficiency ($Q_{\rm abs}\propto \nu^{\beta}$) should be taken into account.
The $\beta$ value is quite uncertain: it approaches zero for very large grain (grey case);
while for small grains, $\beta$ strongly depends on the composition, i.e. 
$\simeq2$ for graphite, $\simeq1$ for silicate, and $\simeq-0.5$ for SiC.
If we adopt $\beta=2$ (see formula 8.16 in Kruegel 2003):
\begin{equation}
  \frac{L_{\rm bol}}{16\pi^2 R^2} \simeq 1.47\times10^{-6}aT^6,
\end{equation}
we get $\tdust=700~K$ and an integrated IR luminosity of
$9.0\times10^{42}$~\lum, calling for \lbol\ of
$3.1\times10^{42}$~\lum and $4.1\times10^{43}$~\lum\ for dust with MRN size distribution 
(Mathis et al. 1977) and an average size of $0.1~\mu$m, respectively.
This can be taken as a lower limit for the dust temperature and required \lbol.

Although with large uncertainties caused by the dust absorption efficiency and grain size,
we can conclude that the derived \lbol\ is basically comparable to the 
observed peak luminosity of PS16dtm, providing a substantial evidence of the 
dust echo associated with the TDE.

\subsection{The rapid emergence of \feii\ lines: metal released from the evaporated dust? }

In the introduction, we have mentioned that various intriguing spectral characteristics
have been found in PS16dtm after its discovery. 
Most remarkably, the optical spectra of PS16dtm show complex emission structures 
with narrow and broad components in the range $\sim4500-5400$~\AA\ arising primarily 
from the \feii\ multiplets. 
Moreover, the \feii\ lines overall become stronger revealed by monitored spectrum
spanning +25 to +189 rest-frame days since the beginning of the rise (MJD 57600). 
The rapid emergence of prominent optical \feii\ lines makes PS16dtm 
resemble a classical NLS1.
The physical mechanism that drives the formation of \feii\ lines in NLS1s is still a 
mystery although it's well known that \feii\ strength is generally correlated with the
Eddington ratio (e.g., Boroson \& Green 1992; Dong et al. 2011).
PS16dtm may offer us a marvellous opportunity to explore the mechanism 
taking advantage of the dynamic transition process from a normal type~1 AGN
to a NLS1 witnessed by us.

To explain its IR light curve, a simple model for PS16dtm is proposed as follows.
After the TDE occurred, the rising bolometric luminosity inevitably evaporated 
the proximate dust gradually until the peak of the flare. 
The new inner radius of the torus recedes to $\sim70$ light days and the original dust 
between 10 to 70 light days is replaced with a gas torus. This part of gas is exposed to 
the radiation field, so it will be ionized and emit emission lines. 
Its velocity is lower than the intermediate-width line
gas since it is more further out ($v\propto r^{-1/2}$),
consistent with the fact that the observed intermediate-width component of \ha\ line
has become narrower while its equivalent width increased (Blanchard et al. 2017).
On the other hand, the metal released from dust will be ionized and 
give pronounced metal lines, particular \feii\ emission lines. 
However, previous studies show that \feii\ emission
cannot be fully explained in the framework of photoionization models 
(Collin \& Joly 2000; Sigut \& Pradhan 2003; Baldwin et al. 2004),
additional mechanisms such as collisional excitation 
might also act (e.g. Joly 1981; V{\'e}ron-Cetty et al. 2006). 
The unbound debris ejected from the disrupted star may produce intense shocks 
by interacting with the gas torus and excite the iron, hence collisional excitation 
is likely also essential here.

A rough estimate of area of \feii\ emitting region can be made by comparing the constant 
temperature collisional excitation model of Joly (1987), which gave an estimate of 
\feii\ flux per unit surface area whatever the detailed heating mechanism.
Using the highest \feii~$\lambda 4570$ intensity ($\rm 3.7\times10^6~erg~s^{-1}~cm^{-2}$) 
in their calculation ($T_{\rm e}=8000$~K, $n_{\rm H}=10^{23}\rm cm^{-1}$) 
and a observed intensity of $\rm \sim10^{-14}~erg~s^{-1}~cm^{-2}$
(see Figure 8 of Blanchard et al. 2017), we can reach a conclusion 
that \feii\ emission area is equivalent to a spherical surface of radius of 
$\sim5\times10^{16}$~cm or $1$ light month. 
This is a lower limit, yet comparable with the size of photon traveling from 
the initial outburst, implying that the physical conditions to account for the
strong \feii\ in PS16dtm is indeed critical.

\subsection{Broad absorption line outflow: driven by radiation pressure?}

While the optical spectra of PS16dtm matches strikingly with NLS1s, its NUV spectra 
taken by \hst/STIS appears very similar to BAL quasars, exhibiting a broad absorption 
trough at around 2700~\AA\ with a velocity width of $\sim10^4$~\kms (Blanchard et al. 2017).  
This feature can be attributed to the blueshifted absorption of \mgii~$\lambda\lambda 2800$ 
doublet and suggests an outflow with velocity of $\gtrsim10^4$~\kms.
Can the outflow be driven by the radiation pressure amplified dramatically by the TDE?

First of all, the radiation pressure the gas experienced becomes higher and higher
along with the rising of the bolometric luminosity.
At the Eddington luminosity, the radiative acceleration is significantly larger than
the gravitational acceleration by a factor of $f$ (100-1000)
depending on the gas column density (Castor et al. 1975; Abbott 1982; Risaliti \& Elvis 2010).
For gas at around the original torus inner radius, that is 10 light days away from
the black hole, its escape velocity ($v_{\rm e}=\sqrt\frac{2G\mbh}{R}$) 
is $\sim10^3$~\kms, and the dynamical time scale is 100-300 days 
($t_{\rm d}=\frac{R}{f^{1/2}v_e}$). Within $t_{\rm d}$, 
the gas can be accelerated to a velocity of $f^{1/2}$ times of the escape velocity,
that is $\gtrsim10^4$~\kms.
This may be sufficient to explain the observed BAL outflow.

Before PS16dtm, only three TDEs have been spectroscopically observed in UV, 
among which PTF15af and iPTF16fnl are X-ray weak and their UV spectra show BALs, 
while ASASSN-14li is X-ray bright and its UV spectrum does not present BALs
(Yang et al. 2017).  PS16dtm shows the same coincidence.
This behaviour is exactly in analogy to quasars with and without BALs.
Despite still with controversy, it is commonly assumed that soft X-rays from 
accretion disk is filtered by a thick shielding gas which prevent over-ionization 
of the outer layer gas and ensure effective radiative acceleration via  
UV resonance absorption lines (Murray et al. 1995).
This may be particularly true for TDE, whose X-ray spectra are generally very
soft, thus a rather moderately thick gas will produce strong absorption.
Further more UV-spectra and X-ray follow-ups of TDEs are necessary to
verify the coincidence or not.

\section{Summary}

Classified as a candidate TDE in a Seyfert~1 galaxy with low-mass BH,
PS16dtm has shown interesting properties in both photometric
and spectral evolutions (Blanchard et al. 2017). 
Surprisingly, we found that the MIR flare was detected 11 days earlier than its optical 
discovery by ASAS-SN, even earlier than Pan-STARRS. 
At the face value, this observation appears to be inconsistent with the 
dust echo interpretation proposed for other TDEs, however, the three unique features 
associated with this event, including close pre-existing dust, high dust covering factor 
and significant noise in very early-time optical light curves due to the host galaxies,  
can indeed provide a satisfactory.
Given the low \mbh\ and \lbol\ before the event, the torus inner radius is only 
$\sim10$ light days.
At the first epoch (E1), the MIR color corresponds to a blackbody temperature
higher than the dust sublimation temperature, implying the dust is located very close 
to the black hole and evaporating. Moreover, at the rising stage,
the MIR luminosity is comparable with the bolometric luminosity, meaning 
the dust covering factor is not low.
At the second and third \wise\ detection epoch (E2 and E3), that is $\sim150$ and 330 
days later, the dust temperature has dropped below the sublimation temperature and
 more clearer evidence for the dust echo is provided.
Particularly, the equilibrium between the dust heating and cooling at E2
results in a bolometric luminosity of $\sim10^{42-45}$~\lum,
depending on the dust absorption efficiency an grain size, 
that is basically consistent with the observed peak luminosity.  
Other possibilities to explain the earlier detection of the flare in MIR than
optical have also been briefly discussed, such as the delay between the optical flare 
and the disruption, the "reprocessing layer" origination of the optical flare. 

In summary, we can still understand the MIR flare of PS16dtm as the dust echo of 
the tidal disruption flare. 
In this case, infrared variability is detected before 
the optical flare, suggesting that IR surveys may trigger follow-up in other band, 
that is important for future infrared survey.
This unique object has also fallen in our sample of nearby galaxies
presenting dramatical MIR flares selected from \wise\ database 
(Jiang et al. in preparation), encouraging us to further discover TDEs using MIR echoes, 
which is expected to be very efficient for those obscured targets. 
 
We have also tried to interpret the newly-emerged prominent \feii\ emission and
\mgii\ BAL under the TDE scenario. As the bolometric luminosity increases,
the inner dust evaporates gradually until to the peak luminosity.
The new torus inner radius has increased to $\sim70$ light days and the region between
10 and 70 light days is replaced with a gas torus. The metal released from dust will 
contribute to the generation of strong \feii\ emission lines. 
In addition to the photoionization, collisional excitation triggered by the interaction
between the tidal debris and gas maybe also be essential for the increasing \feii\ emission. 
On the other hand, the exposed gas can be accelerated by the radiation pressure to a observed
velocity $\sim10^4$~\kms\ (Blanchard et al. 2017) within a dynamical time scale, 
resulting in the observed \mgii\ BAL.
The non-detection of X-ray in PS16dtm is in analogy to BAL quasars, 
in which X-ray is largely absorbed by the shielding gas.

Lastly, we caution that the TDE nature of PS16dtm is not
unquestionable in light of its odd properties, especially the optical light curve
stays persistently at Eddington luminosity for almost half a year instead of 
decaying immediately with time as theory-predicated $t^{-5/3}$ form. 
The MIR flare found in this work
has been interpreted as a TDE echo, but can also an echo of outburst induced 
by other unknown physical processes.
Further follow-up monitorings may help us disentangle the mysteries of this
unique object.
%}

\acknowledgements

We thank the anonymous referee for a thorough report and many constructive
comments that help us improve this work.
We thank Subo~Dong and Ben Shappee for kindly providing us the ASAS-SN data when
we prepared the draft.
We thank Nicholas Stone for his very nice and timely comments on the interpretation 
of the early MIR detection after the draft was posted on arXiv.
This work is supported by the National Basic Research Program of China
(grant No. 2015CB857005), NSFC (NSFC-116203021, NSFC-11421303, NSFC-11733001, NSFC-11303008), 
Joint Research Fund in Astronomy (U1431229, U1531245, U1731104) under cooperative agreement
between the NSFC and the CAS and the Fundamental Research
Funds for the Central Universities.
This research makes use of data products from the
{\emph{Wide-field Infrared Survey Explorer}, which is a joint
project of the University of California, Los Angeles,
and the Jet Propulsion Laboratory/California Institute
of Technology, funded by the National Aeronautics and
Space Administration. This research also makes use of
data products from {\emph{NEOWISE-R}}, which is a project of the
Jet Propulsion Laboratory/California Institute of Technology,
funded by the Planetary Science Division of the
National Aeronautics and Space Administration.

\end{document}